\documentclass[%
 aip,
 amsmath,amssymb,
 reprint,%
]{revtex4-1}
\usepackage{hyperref}
\usepackage{graphicx}
\usepackage[utf8]{inputenc}
\usepackage{amsmath}
\usepackage{mathtools}
\usepackage{algorithm}
\usepackage{algpseudocode}
\usepackage{verbatim}
\usepackage{amssymb}
\usepackage{color}
\usepackage{mathtools}
\usepackage{bm}
\usepackage{stmaryrd}
\usepackage{subfigure}
\usepackage{tikz}
\usepackage{ifthen}
\usepackage{physics}
\usepackage{soul}
\usepackage[normalem]{ulem}
\hypersetup{
    colorlinks=true,
    linkcolor=blue,
    filecolor=gray,
    urlcolor=blue,
    citecolor=blue,
    }
\usetikzlibrary{calc}
\usepackage{cancel}

\definecolor{olivegreen}{HTML}{808000}

\begin{document}

\date{\today}\title{Minimal pole representation for spectral functions}
\author{Lei Zhang}
\affiliation{Department of Physics, University of Michigan, Ann Arbor, Michigan 48109, USA}
\author{Andr\'e Erpenbeck}
 \affiliation{Department of Physics, University of Michigan, Ann Arbor, Michigan 48109, USA}
\author{Yang Yu}
\affiliation{Department of Physics, University of Michigan, Ann Arbor, Michigan 48109, USA}
\author{Emanuel Gull}
\email{egull@umich.edu}
\affiliation{Department of Physics, University of Michigan, Ann Arbor, Michigan 48109, USA}

\begin{abstract}
Representing spectral densities, real-frequency, and real-time Green's functions of continuous systems by a small discrete set of complex poles is an ubiquitous problem in condensed matter physics, with applications ranging from quantum transport simulations to the simulation of strongly correlated electron systems.
This paper introduces a method for obtaining a compact, approximate representation of these functions, based on their parameterization on the real axis and a given approximate precision. We show applications to typical spectral functions and results for structured and unstructured correlation functions of model systems.
\end{abstract}

\maketitle
\section{Introduction}
The accurate representation of a continuous quantum system by a few discrete degrees of freedom is a fundamental problem in condensed matter science. Applications of this problem appear whenever the solution of quantum impurity problems requires a Hamiltonian representation \cite{Caffarel94,Zgid12,Shee19,Zhu19} with a few degrees of freedom \cite{Caffarel94,Koch08,Senechal10,Liebsch12,MejutoZaera20}; whenever transport in open systems is modeled by coupling dissipative bath states to a correlated subsystem \cite{Tanimura89,Arrigoni13,Tanimura20}, and wherever fast quasi-analytic calculations in a Green's function language are needed \cite{Gazizova24}.

Compact and accurate representations of Green's functions are essential in computational many-body simulations \cite{Boehnke11,Kanaenka16,Shinaoka17,Shinaoka18,Li20,Kaye21,Dong22,Kaye22}. A compact and accurate representation respecting the analytic properties of equilibrium Green's function has recently been introduced \cite{zhang2024p1,zhang2024p2} in the context of analytic continuation \cite{Jarrell96,Rothkopf13,Fei21,Fei21_Cara,ying2022analytic,ying2022pole,Huang23} of imaginary-frequency Matsubara Green's functions and is based on the realization that most objects of interest in many-body theory, such as retarded or advanced Green's functions, susceptibilities, and spectral functions are related to Nevanlinna functions \cite{Fei21,Fei21_Cara}. The retarded Green's function in particular is analytic in the upper half of the complex plane, its imaginary part coincides (up to a constant prefactor) with the spectral function just above the real axis, and all of its poles lie in the lower half of the complex plane.  With the help of suitable holomorphic mappings, the problem of accurately approximating the Green's function in the upper half plane  simplifies to an exponential approximation problem~\cite{zhang2024p1,zhang2024p2,ying2022pole,ying2022analytic}, for which well-established signal processing methodology exists \cite{Prony1795,hua1990matrix,roy1989esprit}.

Although functions known directly on the real axis do not suffer from the ill-conditioning of the analytic continuation problem, in practical applications they, too, benefit from systematically improvable compact representations in terms of poles in the complex plane. Such representations enable fast calculations of diagrams in perturbation theory \cite{Gazizova24,Kaye24}, fast propagation of real-time quantities in hierarchical equation of motion calculations \cite{Tanimura89,Arrigoni13,Tanimura20}, and a systematic analysis of the features contained in spectral data \cite{zhang2024p1,zhang2024p2}.

This paper will present systematic methods for constructing approximations of spectral functions from real-frequency spectral data and show applications to real-frequency spectra, as well as their real-time Fourier transforms. The remainder of the paper is organized as follows. In Sec.~\ref{sec:method}, we introduce the minimal pole method (MPM) for constructing compact approximations of spectral functions. Section~\ref{sec:num_res} presents numerical results that demonstrate the performance of the method on representative spectral functions, Green’s functions in the complex plane, and bath correlation functions for both synthetic and structured spectral data. Finally, Sec.~\ref{sec:conclusion} provides a summary and concluding remarks.

\section{Method}\label{sec:method}
The main object of interest in this work is a real-valued spectral function \( A^\text{ex}(\omega) \)  known on the real axis, $\omega \in \mathbb{R}$. 
We aim to find an approximation to this function consisting of $2M$ poles in the complex plane of the form
\begin{equation}\label{eq:approx}
   {A}(z) = \sum_{l = 1}^M \left(\frac{A_l^{(\mathrm{dn})}}{z - \xi_l^{(\mathrm{dn})}} + \frac{A_l^{(\mathrm{up})}}{z - \xi_l^{(\mathrm{up})}}\right) \; ,
\end{equation}
 with $z\in \mathbb{C}$ and ${A}(z=\omega) \approx A^\text{ex}(\omega)$, where  \( A_l^{(\mathrm{dn})} \) and \( \xi_l^{(\mathrm{dn})} \) are complex weights and pole locations  in the lower half of the complex plane, while \( A_l^{(\mathrm{up})} \) and \( \xi_l^{(\mathrm{up})} \) are their counterparts in the upper half-plane. 

Since the spectral function is real, the poles satisfy the following symmetry:
\begin{equation}\label{eq:pole_symmetry}
\xi_l^{(\mathrm{up})} = (\xi_l^{(\mathrm{dn})})^* \;\;\; \text{and} \;\;\; A_l^{(\mathrm{up})} = (A_l^{(\mathrm{dn})})^* \; ,
\end{equation}
and we restrict most of the following discussion to recovering the $M$ poles in the lower half of the complex plane. These poles are related to those of the corresponding retarded Green's function.

\subsection{Minimal Pole Method}\label{sec:mpm}

\begin{figure}[tbh]
    \centering
\includegraphics[width=1.0 \columnwidth]{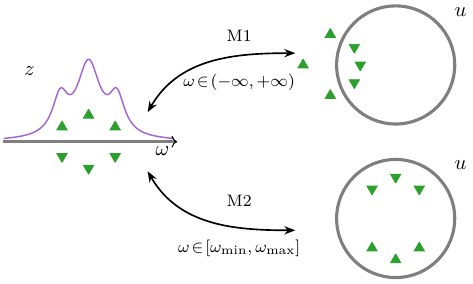}
\caption{Left: The original complex plane \( z \) with the spectral function on the real-frequency axis (purple) and pole locations (green triangles). 
Right: Representations using two different mappings (M1 and M2). 
Top right: The complex plane \( u \) after applying the holomorphic mapping (\ref{eq:con_map1}), which transforms the entire real-frequency range \( (-\infty, +\infty) \) onto the unit circle. This mapping places poles from the lower (upper) half-plane into the interior (exterior) of the unit circle. 
Bottom right: Similar to M1, but employing the holomorphic mapping (\ref{eq:con_map2}). This transformation maps the finite real-frequency interval \([\omega_{\rm min}, \omega_{\rm max}]\) onto the unit circle, positioning all poles inside the unit circle.}\label{fig:con_map}
\end{figure}
As first discussed in the context of analytic continuation in Refs.~\cite{ying2022analytic,ying2022pole}, applying a holomorphic mapping to Matsubara data followed by the application of the residue theorem reformulates the continuation problem into an exponential approximation or `Prony' problem. More recently, Refs.~\cite{zhang2024p1,zhang2024p2} demonstrated that once a continuous imaginary interval is adequately approximated from discrete Matsubara data, this methodology ensures systematic convergence of the spectral function with increasing data precision.
The spectral approximation problem on the real axis can be formulated in much the same way, with the main difference between the analytic continuation and the real-frequency problem being the holomorphic mappings. In this paper, we abbreviate the minimal pole method as MPM.

Among the infinitely many possible choices, we find the following two holomorphic mappings particularly useful for the real-frequency problem: one is used to construct a controlled approximation over the entire real-frequency axis \( \omega \in (-\infty, +\infty) \), while the other is designed for controlled approximation over a finite interval \( \omega \in [\omega_{\min}, \omega_{\max}] \) of interest. We note that other mappings are possible.

For the former case, which takes the spectral weight of \( A(\omega) \) over the entire frequency range as input, we propose applying the M\"{o}bius transform
\begin{align}\label{eq:con_map1}
  \left\{\begin{array}{lll}
    u &= f(z) &= \frac{z + i\omega_p}{z - i\omega_p} \\
    z &= f^{-1} (u) &= i\omega_p\frac{u + 1}{u - 1}
    \end{array}
  \right. ,
\end{align}
where $\omega_p \in \mathbb{R}^+$ is a free parameter, chosen in practice such that the mapped poles are well separated—typically on the same order as the width of the energy window (see Sec.~\ref{sec:num_res} for concrete examples). We denote this mapping as M1.

For the latter case, where \( A(\omega) \) has compact support in the interval \( [\omega_{\rm min}, \omega_{\rm max}] \), we propose applying the transform 
\begin{align}\label{eq:con_map2}
  \left\{\begin{array}{lll}
    u &= f(z) &= z_{\rm s} + \sqrt{z_{\rm s}^2 - 1} \text{ with } z_{\rm s} = \frac{z-\omega_{\rm m}}{\Delta \omega_{\rm h}}  \\
    z &= f^{-1}(u) &= \frac{\Delta \omega_{\rm h}}{2}(u + \frac{1}{u}) + \omega_{\rm m}
  \end{array}
  \right. ,
\end{align}
where $\omega_{\rm m} = (\omega_{\rm min} + \omega_{\rm max}) / 2$, $\Delta \omega_{\rm h} = (\omega_{\rm max} - \omega_{\rm min}) / 2$, and the branch of the square root is chosen such that $|u| \leq 1$. We denote this mapping as M2.
Illustrations of both M1 and M2 are shown in Fig.~\ref{fig:con_map}, with a detailed performance comparison provided in Sec.~\ref{sec:num_res}.

By applying these mappings $u = f(z)$, the real axis or the finite interval are mapped onto the unit circle $\partial \bar{D}$ in the complex $u$-plane and Eq.~(\ref{eq:approx}) transforms into
\begin{equation}
    \!\!{A}'(u) \! = \! \sum_{l = 1} ^M \!\left(\!\frac{A_l'^{\rm (dn)}}{u - \xi_l'^{\rm (dn)}} \! + \! \frac{A_l'^{\rm (up)}}{u - \xi_l'^{\rm (up)}}\!\right) \! + \text{analytic part} ,
\end{equation}
where $A(z) \equiv A'(u)$. Since the function is known on the unit circle of the mapped plane, one may apply the residue theorem to extract the location of the mapped poles via contour integrals along the unit circle. As shown in Refs.~\cite{ying2022analytic,ying2022pole}, after defining the contour integral $h_k$ as
\begin{equation}\label{eq:h_k_def}
    h_k := \frac{1}{2\pi i} \int_{\partial \bar{D}} du \, A'(u) u^k, \quad k \geq 0,
\end{equation}  
the residue theorem implies  
\begin{equation}\label{eq:h_k}
    h_k = \sum_l A_l' \xi_l'^k \; ,
\end{equation}
which is a Prony-type problem (see Appendix~\ref{sec:prony} for a discussion of Prony approximation problems and their solution). Here, $\{A_l', \xi_l'\}$ represent the poles mapped inside the unit circle. As illustrated in Fig.~\ref{fig:con_map}, the poles inside the unit circle correspond to those mapped from the lower half-plane in M1, and to all poles in M2.

For the mapping M1 of Eq.~(\ref{eq:con_map1}), the contour integral Eq.~(\ref{eq:h_k_def}) simplifies to  
\begin{equation}\label{eq:prony_M1}
    h_k = \frac{1}{2\pi} \int_0^{2\pi} d\theta \, A(\omega_p \cot( 
 \theta / 2)) e^{i(k+1)\theta}.
\end{equation}
Similarly, the contour integral Eq.~(\ref{eq:h_k_def}) for M2 of Eq.~(\ref{eq:con_map2}) simplifies to  
\begin{equation}\label{eq:prony_M2}
    h_k = \frac{1}{\pi} \int_0^{\pi} d\theta A(\omega_{\rm m} + \Delta \omega_{\rm h} \cos\theta) \cos(k+1)\theta.
\end{equation}

Given spectral functions on the real axis, the integrals \( h_k \) (\( k = 0, 1, 2, \ldots \)) can be numerically evaluated to high precision using standard quadrature rules.  
Since the poles \( \{A_l', \xi_l'\} \) lie strictly inside the unit circle, \( h_k \) decays with increasing \( k \) according to Eq.~(\ref{eq:h_k}). 
Therefore, the computation of \( h_k \) can be terminated once the target precision \( \varepsilon \) is reached—i.e., when \( |h_k| \leq \varepsilon \)—or when a maximum number of terms \( k_{\rm max} \) is reached if \( h_k \) decays slowly, whichever occurs first.

Knowing $h_k$, the Prony-type approximation problem of Eq.~(\ref{eq:h_k}) can then be solved with standard solution methods for the Prony problem, resulting in $M$ values $A_l'^{\rm (dn)}$ and $x_l'^{\rm (dn)}$. $M$ is either determined a priori as a fixed number of poles, or adjusted dynamically such that a target precision is reached. 
Appendix~\ref{sec:prony} contains more details for the Prony problem and its solution with the ESPRIT method. 

Finally, the values $x_l^{\rm (dn)}$ are recovered using the inverse transform~\cite{ying2022analytic,ying2022pole}
\begin{align}
    \xi_l^{\rm (dn)} & = f^{-1}(\xi_l'^{\rm (dn)}) \notag \\
     & = \left\{\begin{array}{ll}
    i \omega_p \frac{\xi_l'^{\rm (dn)} + 1}{\xi_l'^{\rm (dn)} - 1} & ,\text{for M1}\\
    \frac{\Delta \omega_{\rm h}}{2}(\xi_l'^{\rm (dn)} + \frac{1}{\xi_l'^{\rm (dn)}}) + \omega_{\rm m} \!\!& ,\text{for M2}
    \end{array}\right. ,
\end{align}
and $A_l^{\rm (dn)}$ are obtained using the relation~\cite{zhang2024p1,zhang2024p2}  
\begin{align}
    A_l^{\rm (dn)} & =  \left.\frac{d z}{d u}\right|_{\xi_l'^{\rm (dn)}} \hspace{-6pt} \times A_l'^{\rm (dn)} \notag \\
    & = \left\{\begin{array}{ll}
    -\frac{2i \omega_p}{(\xi_l'^{\rm (dn)} - 1)^2} A_l'^{\rm (dn)} &, \text{ for M1}\\
    \frac{\Delta \omega_{\rm h}}{2}\left(1 - \frac{1}{\xi_l'^{\rm (dn)^2}}\right) A_l'^{\rm (dn)} \!\!&, \text{ for M2}
    \end{array}\right. .
\end{align}
This procedure yields the poles in the lower half of the complex plane. The poles in the upper half-plane are determined using the symmetry relation of Eq.~(\ref{eq:pole_symmetry}).

In the case of multi-orbital systems and matrix-valued retarded Green's functions, it is often physically insightful to approximate the system using shared complex poles with matrix-valued weights. In this case, a matrix-valued generalization of the ESPRIT algorithm~\cite{zhang2024p2} can be employed.

\subsection{Applications}
\subsubsection{From the spectral function to the Green's function on the complex plane}
The Green's function \( G(z) \) in the frequency domain is related to the spectral function \( A(\omega) \) by~\cite{Smit22,negele2018quantum,Mahan13,Stefanucci13,Coleman15}   
\begin{equation}\label{eq:green_gen}
\left\{\begin{array}{lll}
    G(z) = \int_{-\infty}^{+\infty} d\omega \frac{A(\omega)}{z - \omega}, \\
    A(\omega) = -\frac{1}{\pi} {\rm Im}[G(\omega + i0^+)].
\end{array}
  \right.
  \end{equation}
The retarded and the advanced Green's functions on the real axis are related to $G(z)$ as
\begin{equation}
G^{\rm ret/adv} (\omega) = G (\omega \pm i 0^+) \;.
\end{equation}
When evaluated on the Matsubara frequencies $i\omega_n$ with $\omega_n = (2n+1)\pi/\beta$ for fermionic  and $\omega_n = 2n\pi/\beta$ for bosonic Green's functions, $n \in \mathbb{Z}$, the function $G(z)$ corresponds to the Matsubara Green's function:  
\begin{equation}
 G^{\rm Mat}(i \omega_n) = G (i \omega_n)\;.
\end{equation}  
Below, we discuss the relationship between the poles in Eq.~(\ref{eq:approx}) and $G(z)$.

The real and imaginary parts of $G(z)$ are closely related to each other. We divide the discussion here into two cases: First, the case close to the real axis, $z = x \pm i0^+$ and second the general case where ${\rm Im} z \neq 0$ is finite.

The Hilbert transform along the real axis is defined as $\mathcal{H} f(x + iy)= \frac{1}{\pi} \mathcal{P}\int_{-\infty}^{+\infty} dx' \frac{f(x' + iy)}{x - x'}$, where $\mathcal{P}$ denotes the Cauchy principal value. For $z = x \pm i0^+$, using the Sokhotski–Plemelj theorem 
$\frac{1}{x \pm i 0^+} = \mp i \pi \delta(x) + \mathcal{P}(\frac{1}{x})$ and the properties of the Hilbert transform: $\mathcal{H} \delta (x) = \frac{1}{\pi x}$ and $\mathcal{HH}f \equiv -f$, the real and imaginary parts of the retarded/advanced Green's function satisfy ${\rm Im} G(\omega \pm i 0^+) = \pm \mathcal{H} {\rm Re} G(\omega \pm i0^+)$, which are the Kramers-Kronig relations~\cite{de1926theory,kramers1927diffusion,toll1956causality}. For ${\rm Im} z$ away from the real axis, since $\mathcal{H} \frac{|y|}{x^2 + y^2} = \frac{x}{x^2 + y^2}$, it can also be shown that ${\rm Im} G(x \pm iy) = \pm \mathcal{H} {\rm Re} G(x \pm iy)$ for $y>0$. In summary, the real and imaginary parts of the Green's function exhibit different parities in the upper- and lower-half of the complex plane:
\begin{equation}
    {\rm Im} G(z) = \text{sign}({\rm Im} z) \; \mathcal{H} {\rm Re} G(z) \;.
\end{equation}

In addition, due to the symmetry
\begin{equation}\label{eq:green_func_symmetry}
  G(z^*) = {G}(z)^* \; ,
\end{equation}  
it suffices to focus on recovering $G(z)$ in the upper half-plane. Because of the symmetry in Eq.~(\ref{eq:pole_symmetry}), $A(\omega)$ can be reconstructed solely from poles in the lower half-plane. 
Furthermore,
\begin{align}\label{eq:green_ret}
    &2\sum_{l=1}^M \frac{A_l^{(\mathrm{dn})}}{\omega + i 0^+ - \xi_l^{(\mathrm{dn})}} = A(\omega) + i\mathcal{H} A(\omega) \notag \\
    &= -\frac{1}{\pi} {\rm Im} G(\omega + i0^+) + \frac{i}{\pi}{\rm Re} G(\omega + i0^+) \notag \\
    &= \frac{i}{\pi} G(\omega + i0^+) \; .
\end{align}  
Therefore, poles in the lower half-plane provide a natural way to perform the Hilbert transform, and the retarded Green's function can be constructed as  
\begin{equation}\label{eq:green_ret_real}
    G(\omega + i0^+) = -2\pi i \sum_{l=1}^M \frac{A_l^{({\rm dn})}}{\omega + i 0^+ - \xi_l^{({\rm dn})}}.
\end{equation}
Replacing $\omega + i0^+$ with $z$ for ${\rm Im} z > 0$ in Eq.~(\ref{eq:green_ret_real}) gives an approximation of $G(z)$ in the upper half-plane 
\begin{equation}\label{eq:green_upper}
    G(z) = -2\pi i \sum_{l=1}^M \frac{A_l^{({\rm dn})}}{z - \xi_l^{({\rm dn})}}, \quad {\rm Im} z > 0.
\end{equation}

As we will show, $A(\omega)$ can be approximated by complex poles to arbitrary precision over $\omega \in (-\infty, +\infty)$ by utilizing the M1 mapping. By Eq.~(\ref{eq:green_ret_real}), the approximation of $G(z)$ at $z = \omega + i0^+$ ($\omega \in (-\infty, +\infty)$) can also be achieved to arbitrary precision.  
Since both the exact Green's function in Eq.~(\ref{eq:green_gen}) and the approximation in Eq.~(\ref{eq:green_upper}) are analytic for ${\rm Im} \, z > 0$ and their differences on the boundary can be made arbitrarily small, the Green’s function throughout the upper half-plane can be fully described by ${A_l^{({\rm dn})}}$ and $\xi_l^{({\rm dn})}$ with arbitrary precision according to the maximum modulus principle. 
This confirms that the complex pole representation is numerically sufficient to capture all features of $G(z)$, which forms the theoretical foundation for works based on the complex pole representation, such as Refs.~\cite{ying2022analytic,ying2022pole,zhang2024p1,zhang2024p2,goswami2024,huang2025}.

\subsubsection{Decomposition of exponentials for real-time evolution}\label{sec:decompose_exp}

In the simulation of open quantum systems, the bath correlation function (BCF), denoted as $C(t)$, plays a central role as it fully characterizes the influence of the environment on the system of interest. Simulation methods, such as the hierarchical equations of motion (HEOM) \cite{Tanimura89, Tanimura20,Tanimura_Nonperturbative_1990, Ishizaki_Quantum_2005,Hu_Communication_2010,Hu_Pade_2011,Duan_Study_2017,Erpenbeck_Extending_2018, Cui_2019,Xu_Taming_2022,Chen_Universal_2022,Takahashi24}, the hierarchy of pure states~\cite{Suess14}, the pseudomodes approach~\cite{Tamascelli18}, auxiliary master equation methods (AMEA) \cite{Arrigoni13,Dorda14,Werner23}, and the recently developed Quasi-Lindblad theory~\cite{Park24}, rely on its efficient expansion of exponentials,
\begin{align}\label{eq:bcf_def}
    C(t) &= \frac{1}{2\pi} \int_{-\infty}^{+\infty} d \omega J(\omega) \left[\coth\left(\frac{\beta \omega}{2}\right) + 1\right] e^{-i\omega t} \notag \\
    & \approx \sum_{l=1}^M \eta_l e^{-\gamma_l t} \; .
\end{align}
The first identity follows from the fluctuation–dissipation theorem for a system coupled to a bosonic reservoir, where $J(\omega)$ represents the bath spectral density. The second line expresses $C(t)$ as a finite sum of decaying complex exponentials, with $\gamma_l \in \mathbb{C}$ (${\rm Re} [\gamma_l] > 0$) and complex weights $\eta_l \in \mathbb{C}$.

Defining the effective spectral function as  
\begin{equation}
    A(\omega) = J(\omega) \left[\coth\left(\frac{\beta \omega}{2}\right) + 1\right] \; ,
\end{equation}
and decomposing it into a sum of poles as in Eq.~(\ref{eq:approx}), the residue theorem ensures that the approximation in Eq.~(\ref{eq:bcf_def}) can be recovered as  
\begin{align}\label{eq:pole2exp}
    \eta_l = -i A_l^{({\rm dn})}, \quad \gamma_l = i \xi_l^{({\rm dn})} \; .
\end{align}

A similar procedure can be readily applied in the zero-temperature limit. In this case, Eq.~(\ref{eq:bcf_def}) simplifies to
\begin{align}\label{eq:bcf_zeroT}
    C(t) &= \frac{1}{\pi} \int_{0}^{+\infty} d\omega \, J(\omega) e^{-i\omega t} \;.
\end{align}
One can then define the effective spectral function as \( A(\omega) = 2 J(\omega) \) for \( \omega > 0 \), and \( A(\omega) = 0 \) otherwise. The relation in Eq.~(\ref{eq:pole2exp}) remains valid in this limit.

In practice, when using the mapping MPM (M1), $A(\omega)$ is accurate across the entire real-frequency axis, ensuring that $C(t)$ remains accurate across the entire real-time axis. In contrast, when using MPM (M2), since the approximation is only accurate over a finite interval, there exists a time cutoff $t_c$ beyond which the approximation of $C(t)$ becomes less accurate. The cutoff $t_c$ can be identified by adjusting the $L$ parameter in the numerical solution of the problem Eq.~(\ref{eq:h_k}) with ESPRIT (see Appendix~\ref{sec:prony}). 
Let $\hat{C}^{(L_1)}(t)$ and $\hat{C}^{(L_2)}(t)$ denote two approximations of $C(t)$ obtained using different $L$ values. If  
\begin{equation}\label{eq:determine_tc}
    |\hat{C}^{(L_1)}(t) - \hat{C}^{(L_2)}(t)| \leq \varepsilon \;,
\end{equation}  
the approximation is assumed to be accurate at time $t$. Conversely, if the difference is uncontrolled, the approximation is considered unreliable. 
In practice, we use $L_1 = 2N / 5$ and $L_2 = N / 2$ in all our implementations.

\begin{figure*}[tbh]
    \centering
\includegraphics[width=2.0 \columnwidth]{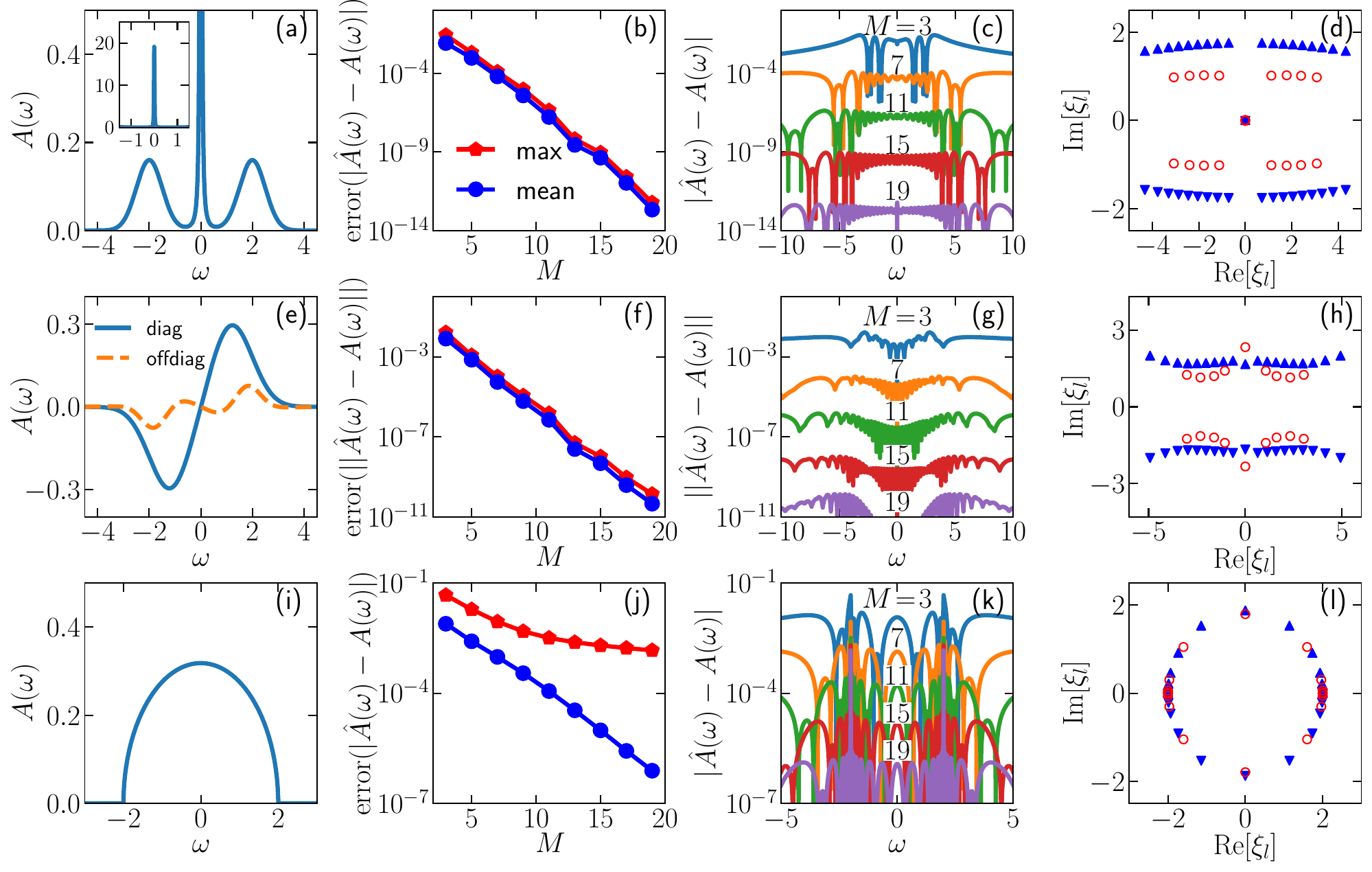}
\caption{Spectral recovery using MPM (M1) for different density of states: Kondo-like (top row), bosonic multi-orbital (middle row), and semicircular (bottom row). 
First column: Exact spectral function \( A(\omega) \).  
Second column: Maximum (red pentagons) and mean (blue circles) of the deviation  as a function of the number of complex poles \( M \) in the lower half-plane.  
Third column: Distributions of the absolute difference as a function of real frequency \( \omega \). 
Fourth column: Distribution of complex poles in the complex plane $z$ for $M=9$ (red circles) and $M=19$ (blue triangles).
}\label{fig:spec_converge}
\end{figure*}

\section{Numerical results}\label{sec:num_res}
\subsection{Approximation of typical spectral functions}\label{sec:num_res1}
We evaluate the pole approximation method by examining three paradigmatic spectral functions that are analytically known on the real axis: a Kondo-like structure, a bosonic multi-orbital spectral function, and a semicircular density of states for MPM (M1). For convenience, we adopt atomic units in Secs.~\ref{sec:num_res1} and \ref{sec:num_res2}, as is customary in studies of model Hamiltonians within the condensed matter community. The free parameter \( \omega_p \) in the mapping is set to \( \omega_p = 2 \). For the calculation of the contour integral, we use \( k_{\rm max} = 3000 \) and a precision of \( \varepsilon = 10^{-12} \), as defined in Sec.~\ref{sec:mpm}.
MPM (M2) yields very similar results, except that the approximation is confined to the finite interval $[\omega_{\rm min}, \omega_{\rm max}]$ (not shown). 

We first examine the performance for a ``Kondo-like'' spectral function of the form  
\begin{equation}\label{eq:spec_kondo}
    A(\omega) = 0.2 g(\omega, -2, 0.5) + 0.6 l(\omega, 0, 0.01) + 0.2 g(\omega, 2, 0.5) \;,
\end{equation}
where $g(\omega, \mu, \sigma) = \frac{1}{\sqrt{2\pi} \sigma} \exp \left\{-\frac{(x-\mu)^2}{2\sigma^2}\right\}$ is a Gaussian function, and  
$l(\omega, \mu, \gamma) = \frac{1}{\pi} \frac{\gamma}{(\omega - \mu)^2 + \gamma^2}$ is a Lorentzian. As shown in Fig.~\ref{fig:spec_converge}(a), the spectrum exhibits both sharp and smooth features but is analytic on the real axis. Fig.~\ref{fig:spec_converge}(b) presents the deviation between the approximated and exact functions as a function of $M$, evaluated on a uniform real-frequency grid spanning $[-10,10]$ with $10^5$ points for concreteness. The approximation remains well controlled even at higher frequencies. As seen in Fig.~\ref{fig:spec_converge}(c), increasing the number of poles leads to exponential convergence of the spectral function to the known solution. Moreover, the sharp peak at the origin is approximated just as accurately as the broader peaks around $\omega = \pm 2$. Finally, Fig.~\ref{fig:spec_converge}(d) shows the distribution of poles, with those corresponding to sharp features located close to the real axis, while those representing smooth features are positioned farther away. 

We further assess the quality of the approximation for a multi-orbital bosonic system, where the spectral functions are given by  $A_{\rm diag}(\omega) = -0.6 g(\omega, -1.2, 0.8) + 0.6 g(\omega, 1.2, 0.8)$ and $A_{\rm off-diag}(\omega) = -0.13 g(\omega, -1.8, 0.5) + 0.1 g(\omega, -1, 1) - 0.1 g(\omega, 1, 1) + 0.13 g(\omega, 1.8, 0.5).$ 
In this case, the absolute difference $|\cdot|$ for scalars is replaced by the matrix norm $||\cdot||$, defined as the maximum absolute difference among all matrix elements. As shown in Figs.~\ref{fig:spec_converge}(f) and (g), the convergence behavior remains similar to the scalar case, with no noticeable slowdown due to the inclusion of off-diagonal elements. The distribution of shared poles, shown in Fig.~\ref{fig:spec_converge}(h), reveals that all poles lie far from the real axis. This is a direct consequence of the absence of sharp spectral features. 

We remark that the above conclusions hold for all of the smooth spectral functions we tested, regardless of their precise form. Given an error tolerance $\varepsilon$, MPM (M1) could always determine a finite set of poles such that $|\hat{A}(\omega) - A(\omega)| \leq \varepsilon$ for any $\omega \in (-\infty, +\infty)$. Furthermore, the number of poles is minimal for a given precision, as ensured by the exponential decay of singular values in the singular value decomposition (SVD) of ESPRIT (see Appendix~\ref{sec:prony}).

Finally, we evaluate the  approximation for a spectral function featuring singularities: a semicircular density of states,  
\begin{equation}\label{eq:spec_semicircle}
    A(\omega) = \left\{
      \begin{array}{cl}
        \frac{1}{2\pi t^2} \sqrt{4t^2 - \omega^2} &, |\omega| \leq 2t \\
        0 &, |\omega| > 2t
      \end{array}
    \right. \; ,
\end{equation}  
which corresponds to the local density of states for an infinite-dimensional Bethe lattice \cite{Metzner89}. The hopping parameter is set to $t = 1$, resulting in square-root singularities at $\omega = \pm 2$.  
As shown in Fig.~\ref{fig:spec_converge}(k), the approximation converges exponentially away from the singularities as $M$ increases. However, as depicted in Fig.~\ref{fig:spec_converge}(j), the presence of singularities slows down convergence.
Despite the challenges posed by these singularities, their impact on practical calculations based on complex poles is typically small, as will be demonstrated in the next section. Fig.~\ref{fig:spec_converge}(l) shows the distribution of poles which cluster around the singular points. 

\subsection{Reconstructing the Green’s function in the complex plane}\label{sec:num_res2}
In this section, we assess the accuracy of the Hilbert transform to recover the Green's function in the entirety of the complex plane for the examples discussed in the previous section.  

\begin{figure*}[tbh]
    \centering
\includegraphics[width=2.0 \columnwidth]{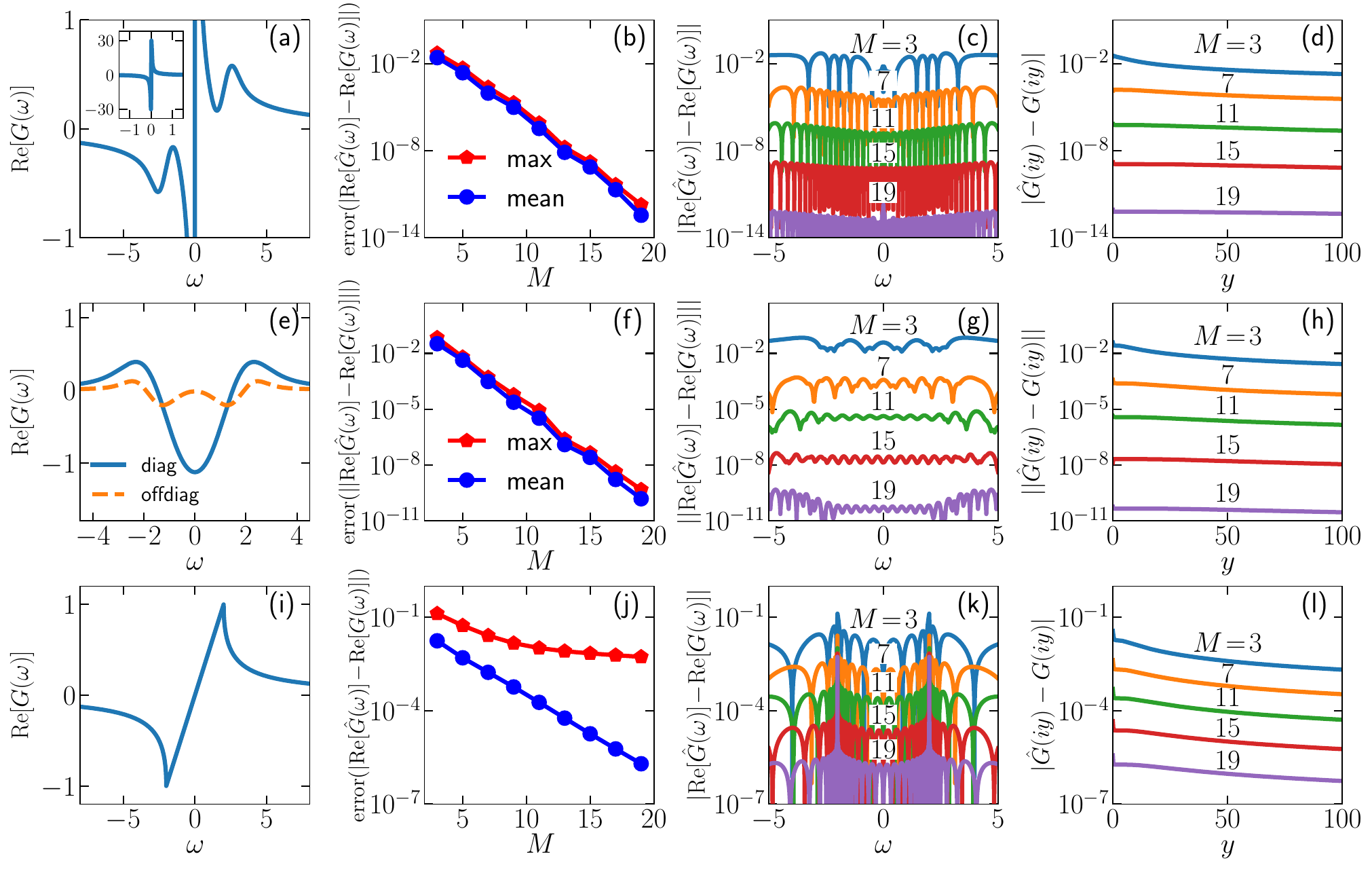}
\caption{Same setup as Fig.~\ref{fig:spec_converge}, but now for the real part of the retarded Green's function and the Matsubara Green's function.  
First column: Exact real part of the retarded Green's function, corresponding to the Hilbert transform of \( \pi A(\omega) \).  
Second column: Error of \( {\rm Re}[G(\omega)] \) as a function of \( M \).  
Third column: Distribution of the absolute difference \( |{\rm Re}[\hat{G}(\omega)] - {\rm Re}G(\omega)| \) as a function of \( \omega \).  
Fourth column: Distribution of the absolute difference \( |\hat{G}(iy) - G(iy)| \) on the positive imaginary axis.
}\label{fig:ret_converge}
\end{figure*}

The exact expression for the real part of the retarded Green’s function corresponding to the Kondo-like spectral function in Eq.~(\ref{eq:spec_kondo}) is given by
\begin{align}
    {\rm Re}[G^{\rm ret}(\omega)] &= 0.2 \pi \tilde{g}(\omega, -2, 0.5) + 0.6 \pi \tilde{l}(\omega, 0, 0.01) \notag \\ 
    & + 0.2 \pi \tilde{g}(\omega, 2, 0.5) \;, \label{eq:DOS_SIC}
\end{align}
where $\tilde{g}(\omega, \mu, \sigma) = \frac{\sqrt{2}}{\pi \sigma} D \left(\frac{\omega - \mu}{\sqrt{2}\sigma}\right)$ and $\tilde{l}(\omega, \mu, \gamma) = \frac{1}{\pi} \frac{\omega - \mu}{(\omega - \mu)^2 + \gamma^2}$ are the Hilbert transforms of the Gaussian and Lorentzian functions, respectively, with $D(x)$ denoting the Dawson function. 
Similar to the spectral function (imaginary part), the real part exhibits sharp features at the origin and smooth variations elsewhere, while remaining analytic on the entire real axis, as shown in Fig.~\ref{fig:ret_converge}(a). The approximation is accurate on all of the real axis, as demonstrated in Figs.~\ref{fig:ret_converge}(b) and (c). Additionally, Fig.~\ref{fig:ret_converge}(d) illustrates the performance of the approximation along the entire positive imaginary axis, which is consistently at least as accurate as on the real axis.

The exact expression for the matrix-valued bosonic spectrum is given by ${\rm Re}[G_{\rm diag}^{\rm ret}(\omega)] = -0.6 \pi \tilde{g}(\omega, -1.2, 0.8) + 0.6 \pi \tilde{g}(\omega, 1.2, 0.8)$ and ${\rm Re}[G_{\rm off-diag}^{\rm ret}(\omega)] = -0.13 \pi \tilde{g}(\omega, -1.8, 0.5) + 0.1 \pi \tilde{g}(\omega, -1, 1) - 0.1 \pi \tilde{g}(\omega, 1, 1) + 0.13 \pi \tilde{g}(\omega, 1.8, 0.5)$. 
As shown in the middle panel of Fig.~\ref{fig:ret_converge}, the convergence behavior closely follows that of the spectrum presented in Fig.~\ref{fig:spec_converge}.

For any smooth spectral function, MPM (M1) determines a minimal number of poles such that, for a given precision $\varepsilon$, the approximation satisfies
\begin{equation}  
    |\hat{G}^{\rm ret}(\omega) - G^{\rm ret}(\omega)| \leq \varepsilon \text{ for any } \omega \in (-\infty, +\infty).
\end{equation}
Since both $G^{\rm ret}(z)$ and $\hat{G}^{\rm ret}(z)$ are analytic in the upper half-plane, the maximum modulus principle ensures that the approximation remains strictly bounded by $\varepsilon$ throughout the entire upper half-plane. 

The analysis continues with the remaining example of semicircular density of states given in Eq.~(\ref{eq:DOS_SIC}), which has the following exact expression for the retarded Green's function: 
\begin{equation}
    {\rm Re}[G^{\rm ret}(\omega)] = \left\{
      \begin{array}{cl}
        \frac{\omega}{2t^2} &, |\omega| \leq 2t \\
        \frac{\omega - {\rm sign}(\omega)\sqrt{\omega^2 - 4t^2}}{2t^2} &, |\omega| > 2t
      \end{array}
    \right. \; .
\end{equation}  
As shown in Fig.~\ref{fig:ret_converge}, the approximation quality closely follows the trend observed in Fig.~\ref{fig:spec_converge}: points far from singularities converge significantly faster than those near them. No computational issues arise along the imaginary axis, as illustrated in Fig.~\ref{fig:ret_converge}(l).

\subsection{Decomposing model spectral densities for real-time evolution}\label{sec:num_res3}
In this section, we evaluate the performance of the methods on the BCF derived from a power-law spectral density with an exponential cutoff. 
To better serve the open quantum systems community, we explicitly provide the units of physical quantities in Secs.~\ref{sec:num_res3} and \ref{sec:struc_spec}.
We compare our results with those obtained using the adaptive Antoulas–Anderson (AAA) algorithm~\cite{Nakatsukasa2018,Nakatsukasa2020}, which operates on the real-frequency domain, as well as with those obtained from the direct decomposition of the BCF using ESPRIT~\cite{roy1989esprit,potts2013parameter}.

The spectral density is defined as  
\begin{equation}
    J(\omega) = \alpha \omega_c^{1-s} \omega^s e^{-\omega / \omega_c} \quad \text{for } \omega \geq 0,
\end{equation}  
and extended to negative frequencies \( \omega < 0 \) by \( J(\omega) = -J(-\omega) \). This form is classified as Ohmic, sub-Ohmic, or super-Ohmic for \( s = 1 \), \( 0 < s < 1 \), and \( s > 1 \), respectively. The corresponding BCF has the analytic form~\cite{brandes2003quantum,weiss2012quantum}
\begin{align}
    C(t) &= \frac{1}{\pi} \alpha \omega_c^{1-s} \beta^{-(s+1)} \Gamma(s + 1) \notag \\
    & \times \left[\zeta \left(s+1, \frac{1 + \beta \omega_c - i \omega_c t}{\beta \omega_c}\right)  
    + \zeta \left(s+1, \frac{1 + i \omega_c t}{\beta \omega_c}\right)\right] ,
\end{align}  
where $\Gamma(z)$ is the gamma function, and $\zeta(z, z')$ is the Hurwitz zeta function.

In this section, we fix the parameters to $\alpha = 1$, $\omega_c = 50 \; {\rm cm}^{-1}\approx 9.4 \times 10^{-3} \; {\rm fs}^{-1}$, and $t_c = 2000 \; {\rm fs}$, following Ref.~\cite{Takahashi24}. The sampling points for ESPRIT are chosen on a uniform grid,  
\begin{equation}\label{eq:t_j_def}
    t_j = \frac{j}{N_t} t_c, \quad (0 \leq j \leq N_t),
\end{equation}  
whereas for AAA, a logarithmic discretization~\cite{walters2017, Takahashi24} over $[-\omega_{\rm max}, \omega_{\rm max}]$ is employed to ensure efficient sampling in the low-frequency domain:  
\begin{equation}
    \omega_j = {\rm sign}(j \! - \! K \! - \! \frac{1}{2}) \omega_{\rm max} \frac{\ln(\frac{K}{K-|j-K-\frac{1}{2}|})}{\ln N_\omega} \;,
\end{equation}  
where $N_\omega = 2K$ is an even integer, and $j$ runs from 1 to $N_\omega$.  

In our simulations, both $N_t$ and $N_\omega$ are fixed at 5000. 
We consistently use $\omega_p = 0.1\; {\rm fs}^{-1} \approx 10.6\; \omega_c$ for MPM (M1), and $\omega_{\rm max} = -\omega_{\rm min} = 0.5\; {\rm fs}^{-1} = 1000\; t_c^{-1} \approx 53.1\; \omega_c$ for MPM (M2), which matches the value used in AAA. 
Additionally, $k_{\rm max}$ is fixed at 1000 for both MPM (M1) and MPM (M2) throughout this section. We note that the choice of $\omega_p$ in MPM (M1) is independent of $t_c$; instead, it is determined by the shape of $A(\omega)$. In contrast, we observe that for such systems, $\omega_{\rm min}$ and $\omega_{\rm max}$ in MPM (M2) are inversely proportional to the time cutoff.

To quantify the performance, we compute the relative errors of the approximations $\hat{A}(\omega)$ and $\hat{C}(t)$ with respect to their exact counterparts:  
\begin{equation}\label{eq:err_Aw}
    \delta A(\omega) = \frac{|\hat{A}(\omega) - A(\omega)|}{|A(0)|}
\end{equation}
and  
\begin{equation}\label{eq:err_Ct}
    \delta C(t) = \frac{|\hat{C}(t) - C(t)|}{|C(0)|}.
\end{equation}  

To analyze convergence as a function of the number of poles $M$, we evaluate the mean error over the approximation interval in the time domain:  
\begin{equation}\label{eq:mean_err_Ct_inside}
    {\rm error}_{\rm inside} = \frac{1}{N_t + 1} \sum_{j = 0}^{N_t} \delta C(t_j).
\end{equation}
To assess the performance on the long-time tail, we compute the mean error over the extended interval $t \in [t_c, nt_c]$:  
\begin{equation}\label{eq:mean_err_Ct_outside}
    {\rm error}_{\rm outside} = \frac{1}{(n-1)N_t + 1} \sum_{j = N_t}^{n N_t} \delta C(t_j),
\end{equation}  
where $t_j$ is defined from Eq.~(\ref{eq:t_j_def}). For concreteness, we set $n = 10$.

\subsubsection{Ohmic bath}
\begin{figure}[tbh]
    \centering
\includegraphics[width=1.0\columnwidth]{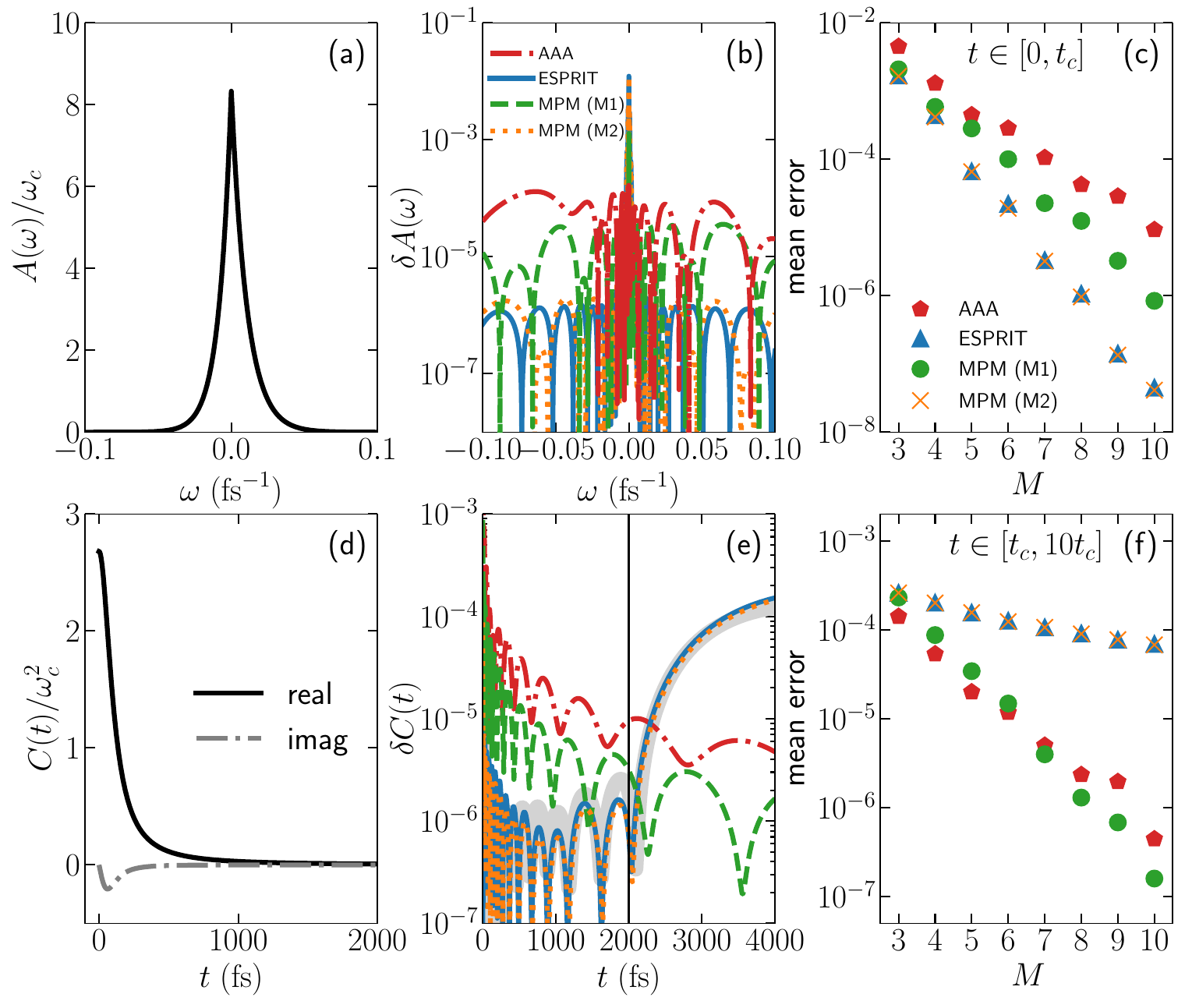}
\caption{
Comparison of AAA (red), ESPRIT (blue), MPM (M1) (green), and MPM (M2) (orange) for an Ohmic bath at $T = 300 \, {\rm K}$ ($\beta \omega_c \approx 0.240$). 
Left panel: (a) Exact effective spectral function $A(\omega)$ and (d) exact bath correlation function $C(\omega)$. 
Middle panel: Relative error of (b) $A(\omega)$ as defined in Eq.~(\ref{eq:err_Aw}) and (e) $C(t)$ as defined in Eq.~(\ref{eq:err_Ct}).  All approximations use $M=8$ terms. The vertical line separates the inside and outside of the approximated real-time interval. The relative difference $ |\hat{C}^{(L_1)}(t) - \hat{C}^{(L_2)}(t)| / |\hat{C}^{(L_1)}(0)|$ is shown by the thick gray curve. 
Right panel: Mean error (Eqs.~(\ref{eq:mean_err_Ct_inside}) and (\ref{eq:mean_err_Ct_outside})) as a function of $M$, evaluated (c) within the interval $[0, t_c]$ and (f) over the long tail $[t_c, 10t_c]$.
}\label{fig:ohmic_300}
\end{figure}

We begin our discussion with the Ohmic bath ($s = 1$) at a temperature of $T = 300 \; {\rm K}$, corresponding to an inverse temperature of $\beta \approx 25.5 \; {\rm fs}$. As a reference, the left column of Fig.~\ref{fig:ohmic_300} displays the exact input function. In the middle panel, we fix $M = 8$ for all methods and compare their performance in both the frequency and time domains.  

In the frequency domain, see Fig.~\ref{fig:ohmic_300}(b), the approximation is more challenging at the origin due to the discontinuity in $dA(\omega)/d\omega$. MPM (M2) and ESPRIT exhibit similar performance, whereas MPM (M1) provides a slightly better approximation at the origin but performs slightly worse away from it. In contrast, AAA approximates the entire interval more uniformly.  

In the time domain, see Fig.~\ref{fig:ohmic_300}(e), within the approximation interval, MPM (M2) and ESPRIT remain nearly identical and provide more accurate results than MPM (M1). However, beyond the interval ($t > t_c$), MPM (M1) retains control over the error, while the errors in MPM (M2) and ESPRIT increase rapidly. AAA is less accurate than MPM (M1) both inside and outside the interval. Here, the frequency cutoff $\omega_{\rm max}$ for MPM (M2) is set to $0.5\; {\rm fs}^{-1}$, determined automatically from Eq.~(\ref{eq:determine_tc}), as indicated by the thick gray line in Fig.~\ref{fig:ohmic_300}(e).  

Finally, the right column of Fig.~\ref{fig:ohmic_300} shows the convergence of the mean error as a function of the number of poles $M$. Within the interval, MPM (M2) and ESPRIT overlap and provide the best error control, while MPM (M1) converges slightly more slowly, and AAA converges the slowest. However, for the long tail, MPM (M2) and ESPRIT exhibit the least control, whereas MPM (M1) and AAA converge significantly faster.

\begin{figure}[tbh]
    \centering
\includegraphics[width=1.0\columnwidth]{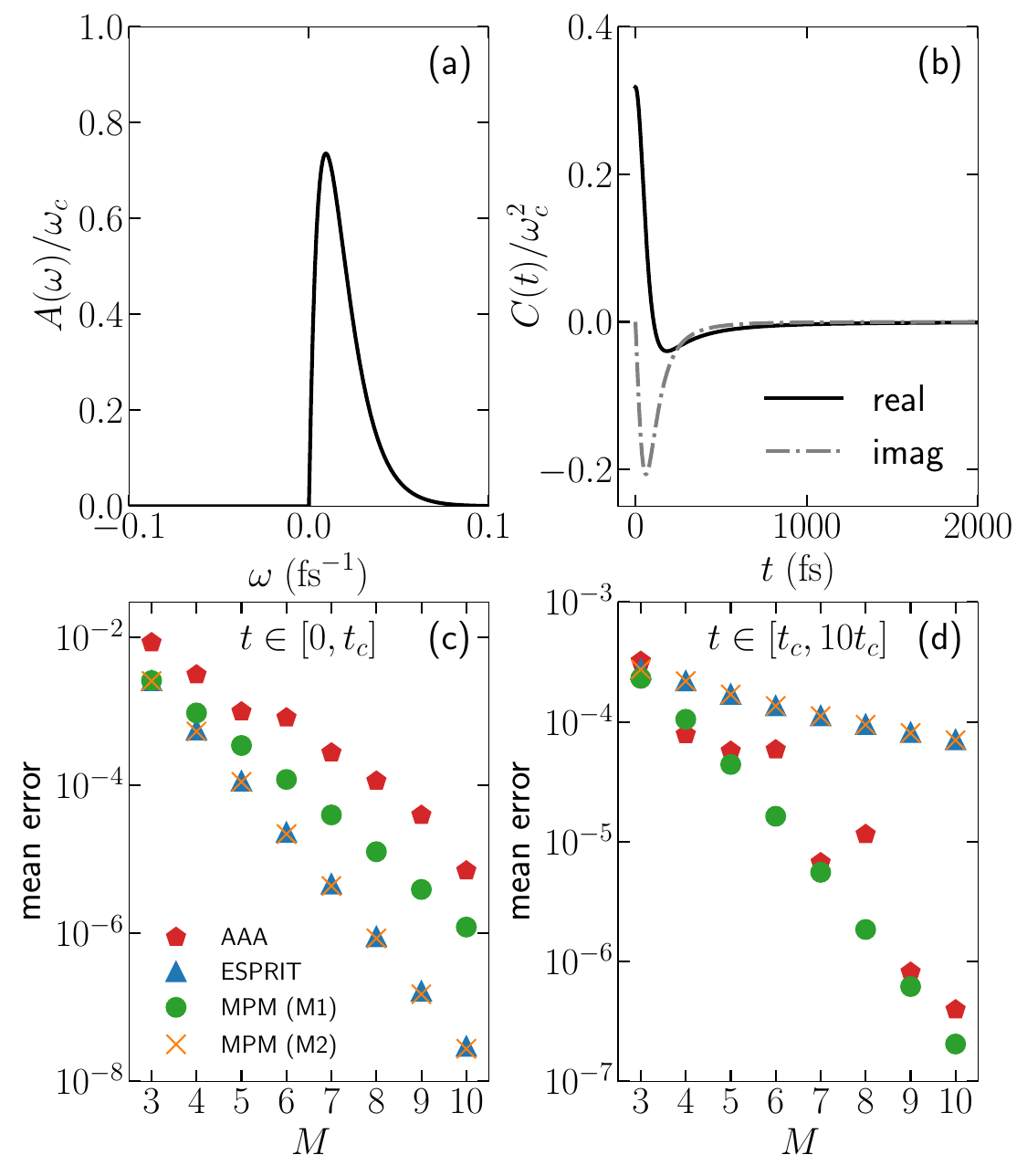}
\caption{Same setup as Fig.~\ref{fig:ohmic_300}, except that temperature is now set to $T = 0.001 \, {\rm K}$ ($\beta \omega_c \approx 7.19\times 10^4$). 
Top panel: (a) Exact spectral function. (b) Exact bath correlation function.  
Bottom panel: Mean error for (c) the approximated interval and (d) the long-tail.
}\label{fig:ohmic_001}
\end{figure}

We now examine the same system at a much lower temperature, $T = 0.001 \; {\rm K}$, corresponding to an inverse temperature of $\beta \approx 7.64 \times 10^{6} \; {\rm fs}$. In this regime, the effective spectrum varies rapidly near the origin. As shown in Fig.~\ref{fig:ohmic_001}, no convergence slowdown with respect to $M$ is observed at low temperature for the methods discussed in this paper. Their performance trends remain consistent with those observed at higher temperatures. Although AAA occasionally achieves an accuracy in the long tail similar to MPM (M1), its convergence behavior is less robust and its accuracy within the finite interval is consistently lower than that of MPM (M1).

\subsubsection{Sub-Ohmic bath}
\begin{figure}[tbh]
    \centering
\includegraphics[width=1.0\columnwidth]{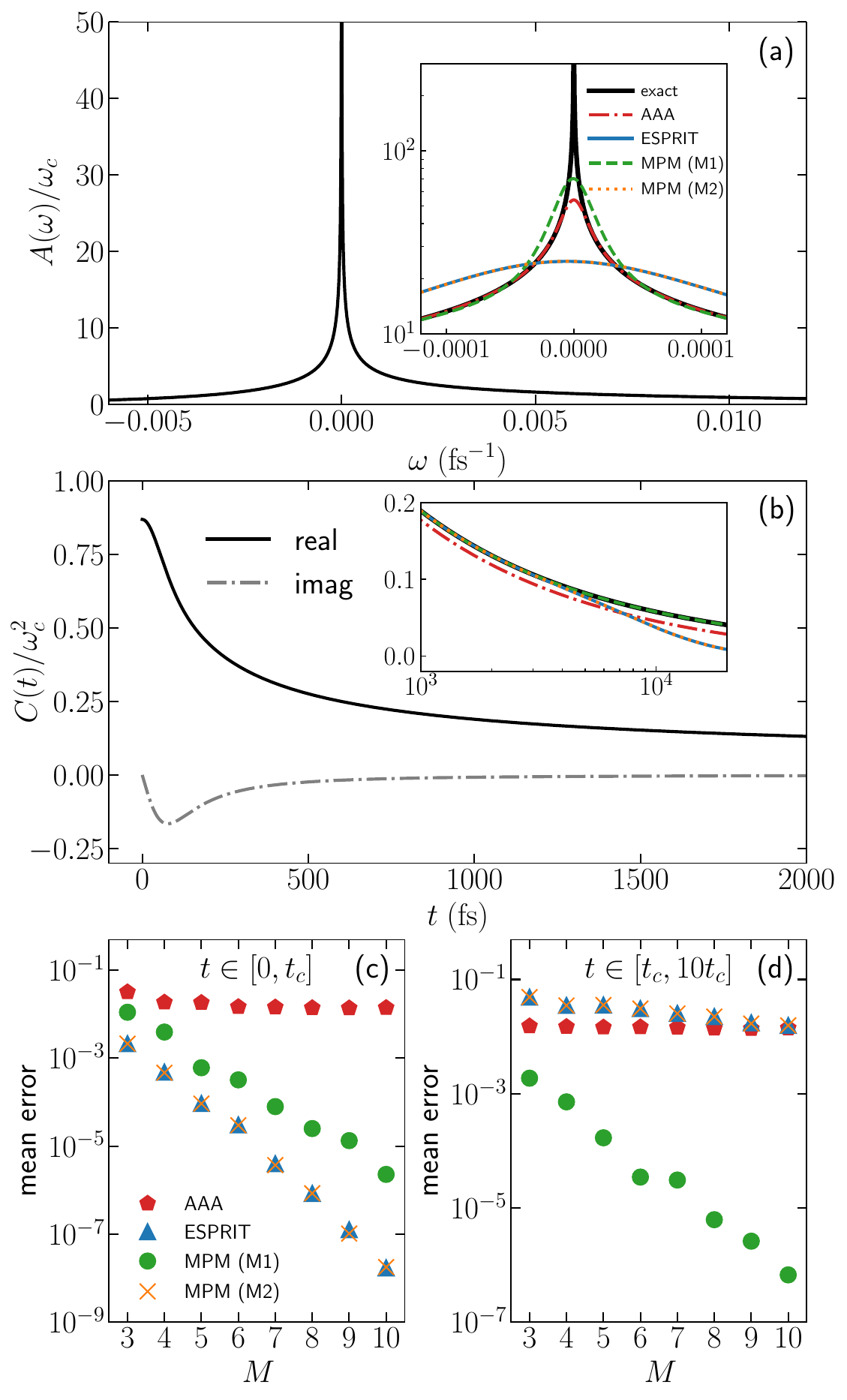}
\caption{Comparison for a sub-Ohmic bath at $T = 50 \, {\rm K}$ ($\beta \omega_c \approx 1.44$). Eight terms are used for all approximations.  
(a) Exact spectral function, which has a divergence at $\omega = 0$.  
(b) Exact bath correlation function, which exhibits an extremely slowly decaying tail for the real part. The insets show the performance of different methods in a zoomed-in region.  
Mean error for the approximated interval and the long tail are shown in subfigures (c) and (d), respectively.
}\label{fig:subohmic_50}
\end{figure}

In this section, we examine a sub-Ohmic bath ($s = 0.5$) at $T = 50 \; {\rm K}$ ($\beta \approx 153 \; {\rm fs}$). The spectral function, which is shown in Fig.~\ref{fig:subohmic_50}(a), diverges at $\omega = 0$, but its integral remains finite, ensuring the existence of the BCF. However, the real part of $C(t)$, which is shown in Fig.~\ref{fig:subohmic_50}(b), exhibits an extremely slowly decaying tail. 

Both AAA and ESPRIT encounter significant issues in this case. As shown in Fig.~\ref{fig:subohmic_50}(c) and (d), AAA completely fails. This failure arises because AAA attempts to approximate the spectral shape on the given grid, causing it to consistently underestimate the exact spectrum near the divergence point, as illustrated in the inset of Fig.~\ref{fig:subohmic_50}(a).  

A similar issue arises in the real-time domain for ESPRIT. Due to the extremely slow decay of $C(t)$, ESPRIT loses error control immediately beyond the finite interval. This presents a fundamental challenge: to maintain error control within $10^{-3}$ over the long tail, for example, $t_c$ would need to be increased beyond $10^7 \; {\rm fs}$. However, since computational constraints limit the number of sampling points to a few thousand, the resulting coarse time grid spacing prevents accurate resolution of the peak in the imaginary part of $C(t)$. 

In contrast, MPM does not suffer from these issues. 
On one hand, MPM (M2) achieves an accuracy comparable to ESPRIT, both within and beyond the approximation interval.
On the other hand, although MPM (M1) converges slightly more slowly than MPM (M2) and ESPRIT within the finite interval, it maintains control over the long tail. In particular, MPM (M1) is the only method that effectively handles the long tail. The success of MPM stems from its approach of approximating spectral moments rather than the spectral shape. Although the spectral function diverges at the origin, its moments remain finite.

\subsection{Decomposing structured spectra for real-time evolution}\label{sec:struc_spec}
\begin{figure}[tbh]
    \centering
\includegraphics[width=1.0\columnwidth]{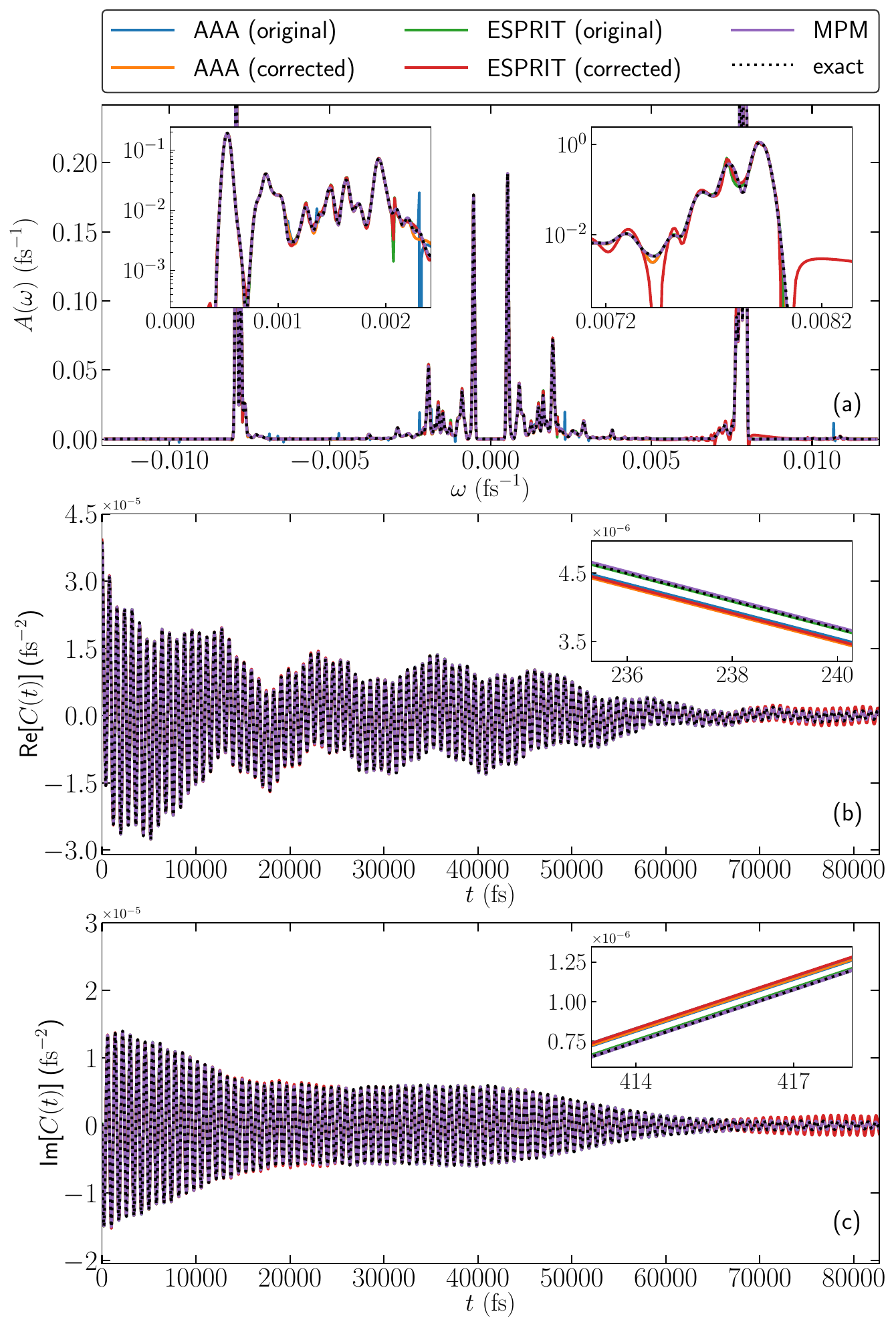}
\caption{Comparison of different approximation methods (solid lines) with the exact results (dashed lines) for a structured spectral function at $T = 300\,{\rm K}$.  
Results are obtained for $M=100$.  
(a) Structured spectrum, (b) real part, and (c) imaginary part of the corresponding bath correlation function.  
The insets show the performance of different methods in magnified regions.}\label{fig:structured}
\end{figure}

Finally, we evaluate the performance of the methods on a structured spectrum derived from exciton--phonon couplings in quantum dots~\cite{Jasrasaria2023, Jasrasaria2021}, which plays a key role in understanding nonradiative exciton dynamics in these systems.

The input data consists of the spectral density $J(\omega)$ sampled on a dense uniform grid. To enable flexible sampling at arbitrary frequency points, the AAA algorithm is applied as a preprocessing step to the effective spectrum at $T = 300\; {\rm K}$, with accuracy prioritized over the number of poles. The error tolerance is set to $10^{-8}$, resulting in 704 poles $\{A_l^{({\rm AAA})}, \xi_l^{({\rm AAA})}\}$, evenly distributed between the upper and lower half-planes. These poles are treated as the exact solution and serve as the reference for all methods compared in this section. In this case, the contour integral in Eq.~(\ref{eq:h_k_def}) for MPM can be evaluated analytically using \( h_k = \sum_{l} A_l^{'({\rm AAA})} (\xi_l^{'({\rm AAA})})^k \) [Eq.~(\ref{eq:h_k})] instead of a numerical integral, thereby improving computational efficiency.

We present the results in Fig.~\ref{fig:structured}, where $M$ is fixed at 100. For AAA, the number of poles to be recovered is set to $2M$, as it captures poles across the entire complex plane. For ESPRIT, the parameters $t_c = 20000\; {\rm eV}^{-1} \approx 8.27 \times 10^4 \; {\rm fs}$ and $N_t=5000$ are used. Issues arise for both the AAA and ESPRIT algorithms. On one hand, the original AAA method consistently recovers some poles on the real axis, leading to non-causal jumps in the spectrum. Denoting these poles as $\{A_l^{(\rm real)}, \xi_l^{(\rm real)}\}$ and their count as $M_{\rm real}$, these spurious poles introduce a constant term in $C(t)$, degrading accuracy in both the real-frequency and real-time domains. To ensure a reliable real-time evolution, the method should be modified to remove these poles. However, as shown in Fig.~\ref{fig:structured}, even after correction, both the approximated spectrum and the BCF exhibit some deviations from the exact results. On the other hand, ESPRIT becomes unstable for some values of the parameter $L$. Although this issue can be mitigated by carefully tuning $L$, allowing ESPRIT to approximate the BCF similarly to MPM, we note that it consistently recovers some poles in the upper half plane. This leads to a divergent amplitude of the BCF at long times, even though the exact BCF exhibits decay. 
To eliminate the asymptotic instability, these non-physical poles must be removed~\cite{dunn2019removing}. In practice, this is done by first obtaining $M$ poles from ESPRIT, discarding the unphysical ones, and then performing a least-squares fit to determine the weights of the remaining poles. Unsurprisingly, we find that once some poles are disregarded, ESPRIT immediately loses error control, as evidenced by its deviation from the exact results in Figs.~\ref{fig:structured}(b) and (c).  

In contrast, MPM does not suffer from these issues. Its poles are automatically placed in the lower half-plane, ensuring that the approximation remains stable and well-controlled. As a result, MPM provides the most accurate approximation in both the time and frequency domains. Furthermore, in this case, MPM (M1) and MPM (M2) yield nearly identical results due to the rapid decay of the long tails in $A(\omega)$. Consequently, we present only MPM (M1) results throughout this section, with $\omega_p$ fixed at $0.04 \; {\rm eV} \approx 9.67 \times 10^{-3}\; {\rm fs}^{-1}$ and $k_{\rm max}$ set to 3000.

\begin{figure}[tb]
    \centering
\includegraphics[width=1.0\columnwidth]{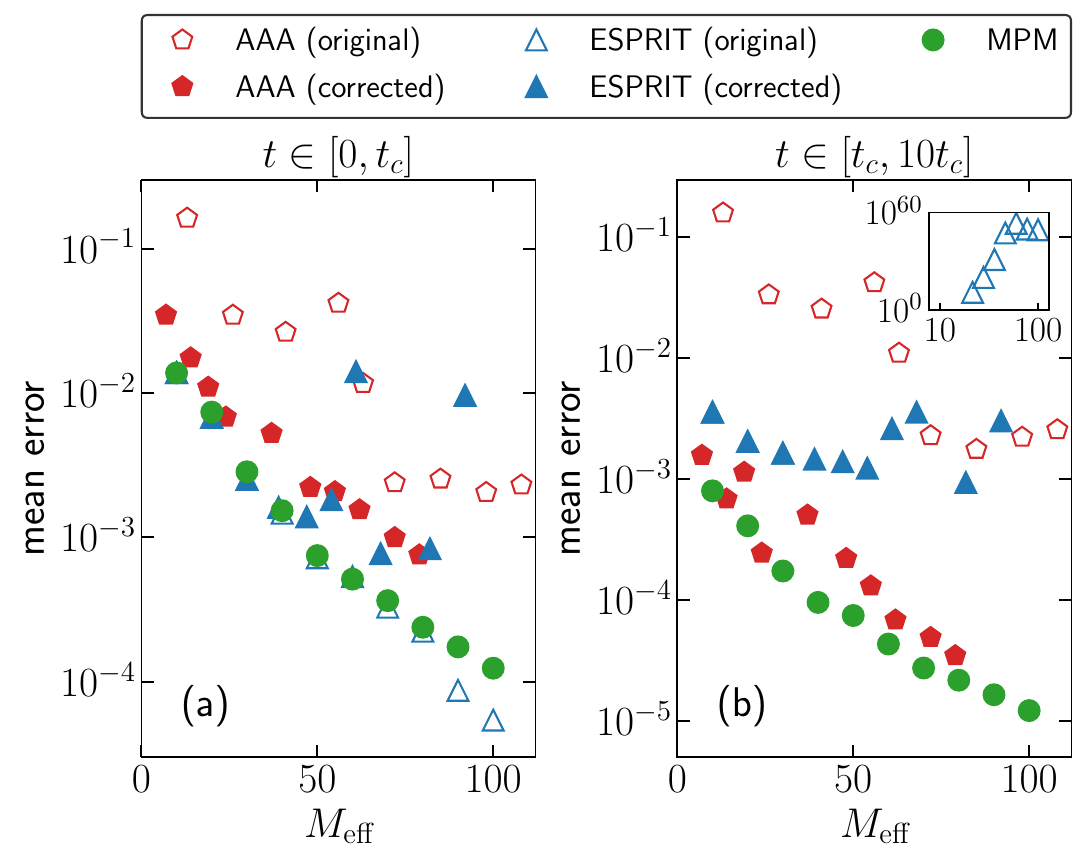}
\caption{Mean error from Eqs.~(\ref{eq:mean_err_Ct_inside}) and (\ref{eq:mean_err_Ct_outside}) as a function of the effective number of poles used in the computation of $C(t)$. 
Unfilled markers represent the original results that should not be used, while filled markers represent the results after correction.  
The inset shows the performance of ESPRIT (original) on a larger scale.}\label{fig:structured_err}
\end{figure}

Fig.~\ref{fig:structured_err} presents the results for $M$ ranging from 10 to 100. To ensure a fair comparison, we use the effective number of poles, $M_{\rm eff}$, defined as the number of poles contributing to the BCF calculation. Specifically, $M_{\rm eff}$ is given by  $M + \frac{M_{\rm real}}{2}$ for the original AAA, $M - \frac{M_{\rm real}}{2}$ for the corrected AAA, $M$ for MPM and the original ESPRIT, and potentially less than $M$ for the corrected ESPRIT if non-physical poles appear. As discussed previously, while the original ESPRIT can sometimes match or even slightly outperform MPM, it frequently produces non-physical poles. Removing these poles leads to a complete loss of error control. For AAA, eliminating real-axis poles consistently improves accuracy both within and beyond the approximation interval. However, AAA remains less robust and less accurate than MPM, reinforcing MPM as the most reliable method among those compared.

\section{Conclusions}\label{sec:conclusion}
In conclusion, we have presented a method to approximate real-frequency spectral functions as a sum of poles in the complex plane. The method is systematically improvable in the sense that, for a given set of real-frequency data and a given approximation precision, a minimal number of poles yielding the most accurate representation possible is generated.

We have shown the behavior of the method for synthetic spectral functions as well as for spectral functions with a definite structure.

We expect that the methodology will find applications whenever a Green's function of a continuous quantum system needs to be represented by a finite (possibly very small) number of discrete degrees of freedom, and 
we have provided examples that are typical applications of the hierarchical equations of motion method.

Our paper is accompanied by an open-source software implementation of the proposed method, written in the programming language {\small PYTHON}~\cite{zhang_2025_15121302,MiniPoleCode}.

\begin{acknowledgments}
We thank Michael Thoss, Samuel Rudge, Salvatore Gatto, Eran Rabani, Bokang Hou and James P. F. LeBlanc for helpful discussions. This material is based upon work supported by the National Science Foundation under Grant No.~2310182. Real-time aspects have been supported by the U.S. Department of Energy, Office of Science, Office of Advanced Scientific Computing Research and Office of Basic Energy Sciences, Scientific Discovery through Advanced Computing (SciDAC) program under Award No. DE-SC0022088.
\end{acknowledgments}

\section*{DATA AVAILABILITY}
The data and the code that support the findings of this study are available from the corresponding author upon reasonable request.

\bibliography{reference}

\appendix
\section{Prony-like Problem and ESPRIT Algorithm} \label{sec:prony}
The extraction of the location of the poles relies on the solution of a Prony-like approximation problem. We therefore briefly summarize the problem and its solution method here.

Given \( N \) values of a function \( f(t) \) sampled uniformly over an interval \([a, b]\) at sampling points $t_j$ and a target precision \( \varepsilon > 0 \), the Prony approximation constructs an approximation of the function as a sum of complex exponentials:  
\begin{equation}\label{eq:prony_problem}
    \left|f(t_j) - \sum_{l=1}^M \eta_l e^{-\gamma_l t_j}\right| \leq \varepsilon, \quad \text{for any } 0 \leq j \leq N-1 \; ,
\end{equation}  
where \( \eta_l, \gamma_l \in \mathbb{C} \), and the sampling points are given by  $t_j = a + j \Delta t$, with $\Delta t = \frac{b - a}{N - 1}$. 

This equation can be reformulated as  
\begin{equation}\label{eq:prony_problem_reformulate}
    \left|f(t_j) - \sum_{l=1}^M R_l z_l^j\right| \leq \varepsilon, \quad \text{for any } 0 \leq j \leq N-1 \; ,
\end{equation}  
where the weights and nodes are defined as  
$R_l = \eta_l e^{-\gamma_l a}$ and $z_l = e^{-\gamma_l \Delta t}$, respectively.

Many numerical methods have been proposed to estimate \( M \), \( R_l \), and \( z_l \) in Eq.~(\ref{eq:prony_problem_reformulate}). These include the Prony approximation method~\cite{beylkin2005approximation, beylkin2010approximation}, the Matrix Pencil Method~\cite{hua1990matrix,sarkar1995using}, and the Estimation of Signal Parameters via Rotational Invariance Techniques (ESPRIT)~\cite{roy1989esprit,potts2013parameter}. Due to its superior performance~\cite{Takahashi24}, we focus exclusively on the ESPRIT algorithm in this paper.

The ESPRIT algorithm utilizes singular value decomposition (SVD) on an \((N-L) \times (L + 1)\) Hankel matrix  
\begin{equation}\label{eq:hankel_matrix}
    H =
    \begin{pmatrix}
        f(t_0)       & f(t_1)     & \cdots & f(t_L) \\
        f(t_1)       & f(t_2)     & \cdots & f(t_{L+1}) \\
        \vdots    & \vdots  & \ddots & \vdots \\
        f(t_{N-L-1}) & f(t_{N-L}) & \cdots & f(t_{N-1})
    \end{pmatrix}
\end{equation} 
expressed as  
\begin{equation}
    H = U \Sigma W \;,
\end{equation}  
where \( U \) and \( W \) are unitary matrices of dimensions \((N-L) \times (N-L)\) and \((L+1) \times (L+1)\), respectively. The matrix \( \Sigma \) is a rectangular diagonal matrix of size \((N-L) \times (L+1)\), where diagonal elements are ordered as \( \sigma_1 \geq \sigma_2 \geq \cdots \geq \sigma_{L+1} \geq 0 \).  In practice, \( L \) is typically chosen between \( N/3 \) and \( N/2 \) to minimize variance~\cite{sarkar1995using}. Throughout our implementation, we set \( L = 2N/5 \) unless otherwise stated.

Given a target tolerance \( \varepsilon \), the number of exponentials \( M \) is estimated as the smallest index satisfying \( \sigma_{M+1} \leq \varepsilon \). This truncation ensures that the minimal number of exponentials is obtained. Next, the nodes \( z_i \) are determined as the eigenvalues of the matrix  
\begin{equation}
    F = (W_0^T)^+ W_1^T \; ,
\end{equation}  
where \( T \) denotes the transpose, \( + \) represents the pseudo-inverse, and \( W_0 \) and \( W_1 \) are extracted from the matrix \( W \) as  
\begin{equation}\label{eq:esprit_Wm}
    W_s = W(1:M, 1+s:L+s), \quad s = 0,1,
\end{equation}  
where both the row and column indices run from \( 1 \) to \( L + 1 \).  
Finally, the weights \( R_l \) are obtained by solving the overdetermined least-squares Vandermonde system 
\begin{equation}\label{eq:solve_weights_matrix}
\begin{pmatrix}
    f(t_0) \\
    f(t_1) \\
    \vdots \\
    f(t_{N-1})
\end{pmatrix}
\!=\!  
\begin{pmatrix}
1 & 1 & \cdots & 1 \\
z_1 & z_2 & \cdots & z_M \\
\vdots    & \vdots    & \ddots & \vdots \\
z_1^{N-1} & z_2^{N-1} & \cdots & z_M^{N-1}
\end{pmatrix}\!\!
\begin{pmatrix}
    R_1 \\
    R_2 \\
    \vdots \\
    R_{M}
\end{pmatrix} .
\end{equation}
Thus, all sampled data are approximated within the tolerance \( \varepsilon \) using a minimal number of exponentials. 

In cases where the sampled function is matrix-valued, a generalization of ESPRIT to matrix-valued functions~\cite{zhang2024p2} can be applied.

\end{document}